\documentclass{elsart}

\usepackage{natbib}
\usepackage{amssymb}
\usepackage{graphicx}

\begin{document}

\begin{frontmatter}

\title{ A Robust Estimation of the Exponent Function in the Gompertz Law}

\author{V. Ibarra-Junquera},\ead{vrani@ipicyt.edu.mx}$\,$
\author
{M.P. Monsivais},\ead{monsivais@ipicyt.edu.mx} $\ \ $
\author
{H.C. Rosu}\ead{hcr@ipicyt.edu.mx}$\ $ \&
\author
{R. L\'opez-Sandoval}\ead{sandov@ipicyt.edu.mx}

\address
{IPICyT - Instituto Potosino de Investigaci\'on Cient\'{\i}fica y
Tecnol\'ogica, \\ \footnotesize Apdo Postal 3-74 Tangamanga, 78231
San Luis Potos\'{\i}, M\'exico}

\begin{abstract}
The estimation of the solution of a system of two differential
equations introduced by Norton {\em et al.} (1976) that is
equivalent to the famous Gompertz growth law is performed by means
of the recent adaptive scheme of Besan\c{c}on and collaborators
(2004). Results of computer simulations illustrate the robustness of
the approach.
\end{abstract}

\begin{keyword}
Gompertz law \sep adaptive scheme \sep diffeomorphism

\medskip

{\em physics/0411088}: $\quad$ version accepted at Physica A
\end{keyword}

\end{frontmatter}


\section{Gompertz growth functions}

Mathematically speaking, the Gompertz law refers to the class of
functions having exponentially decreasing logarithmic derivatives.
It has been introduced in 1825 by B. Gompertz in a paper 
on the law of human mortality \cite{Gomp1825}. He noted that for
people between 40 and 100 years ``{\em the rate of mortality
variable with age measures a force of death which grows each year by
a fraction, always the same, of its present value}''. According to
Winsor \cite{Win}, the possible application of the Gompertz curve in
biology was first spelled out in 1926 by Sewall Wright in a few
remarkable statements: {\em`` In organisms, on the other hand, the
damping off of growth depends more on internal changes in the cells
themselves ... The average growth power as measured by the
percentage rate of increase tends to fall at a more or less uniform
percentage rate, leading to asymmetrical types of s-shaped
curves...''.}

If the size $z(t)$ of a growing structure evolves according to the
equation \cite{js03}
\begin{equation}\label{Gomp0}
\dot{z}=z\ln \left(\frac{S}z\right)~,
\end{equation}
we say that its growth is of Gompertz type. The evolution is
continuous from a given initial stage to a plateau value $S$. In a
Nature letter on the growth of tumours, Norton {\em et al}
\cite{nor} wrote the Gompertz law as the system of the following two
first order differential equations
\begin{eqnarray}
\dot{Z}_1 &=& K_1 Z_1 Z_2 \label{Ec1}\\
\dot{Z}_2 &=& -K_2 Z_2~, \label{Ec2}
\end{eqnarray}

\noindent where $Z=(Z_1,Z_2)\in \mathbb{R}^2$, $K_i>0$, $Z_1$ is the
volume of the tumour at time $t$, and $Z_2$ is a function entirely
described by the second equation (\ref{Ec2}) that gives the
difference in growth with respect to a pure exponential law.
According to Norton, $K_2 Z_2$ gives the fraction of the volume that
doubles in size during the instant $dt$. Thus, $Z_2$  that we call
for obvious reasons the {\em Gompertzian exponent function} is of
special interest  and we would like to determine it with high
accuracy even though we know neither the initial conditions
 for $Z_1$ and $Z_2$ nor $K_2$.
Norton {\em et al} wrote the solution of the system in the following
form

\begin{eqnarray}
{Z}_1(t) &=& {Z}_1(0) \exp \left[ \left( \frac{K_1}{K_2}
\right){Z}_2(0) \left\{ 1-\exp(-K_2 t)  \right\}  \right]~,
\label{sol1}\\
{Z}_2(t) &=&   \left( \frac{K_2}{K_1} \right) \ln \left[ \frac{S}
{{Z}_1(t)} \right]~.\label{sol2}
\end{eqnarray}

We will treat $Z_1$ and $Z_2$ as states of a dynamical system that
in our case is the evolution of a tumour. The fundamental concept of
state of a system or process could have many different empirical
meanings in biology and in our case the first state $Z_1$ is just
the size of the tumor whereas $Z_2$ is the deviation of the growth
rate from the pure exponential growth. In general terms, a
potentially useful tool in biology is the reconstruction of some
specific states under conditions of limited information. For animal
tumors, it is not trivial to know their state $Z_1$ at the initial
moment and most often we do not know the instant of nucleation that
can be determined only by extrapolation of the fitting to the
analytic solutions of growth models, such as Eqs.~(\ref{sol1}) and
(\ref{sol2}). In this paper, the main goal is to show that an
excellent alternative procedure for estimating the phenomenological
quantities of the tumor growing process in the frequent case in
which we do not know the initial conditions and the parameter $K_2$
is the recent adaptive scheme for state estimation proposed by
Besan\c con and collaborators \cite{Besancon 2004}. In addition,
what is generally measured, i.e., the output $y$, is a function of
states that we denote by $h(Z)$ and in the particular case of tumors
one usually measures their volume. Then:
\begin{eqnarray}
y=h(Z)=Z_1~.
\end{eqnarray}
The mathematical formalism of the adaptive scheme that follows
relies entirely on the Lie derivatives of the function $h(Z)$ that
are defined in the next section. By a Lie mapping, we are able to
write the Gompertz-Norton system in Besan\c con's matrix form
(system $\digamma$ below) that allows to write the corresponding
adaptive algorithm (the dynamical system $\hat{\digamma}$ and its
explicit Gompertz form  $\hat{\digamma} _G$ below).

\section{The adaptive scheme}

Taking into account the fact that rarely one can have a sensor on
every state variable, and some form of reconstruction from the
available measured output data is needed, an algorithm can be
constructed using the mathematical model of the process to obtain an
estimate, say $\hat{X}$ of the true state $X$. This estimate can
then be used as a substitute for the unknown state $X$. Ever since
the original work by Luenberger \cite{Luenberger 1966}, the use of
state `observers' has proven useful in process monitoring and for
many other tasks. The engineering concept of observer means an
algorithm capable of giving a reasonable estimation of the
unmeasured variables of a process using only the measurable output.
Even more useful are the so-called {\em adaptive} schemes that mean
observers that are able to provide an estimate of the state despite
uncertainties in the parameters.
The so-called high gain techniques proved to be very efficient for
state estimation, leading in control theory to the well-known
concept of \emph{high gain observer} \cite{J.P.Gauthier92}. The gain
is the amount of increase in error in the observer's structure. This
amount is directly related to the velocity with which the observer
recovers the unknown signal. The high-gain observer is an algorithm
in which the amount of increase in error is {\em constant} and
usually of high values in order to achieve a fast recover of the
unmeasurable states.
In case of dynamical systems depending on unknown parameters, the
design of the observer has to be modified appropriately in order
that the state variables and parameters could be estimated. This
leads to the so called \emph{adaptive observers}, i.e., observers
that can change in order to work better or provide more fit for a
particular purpose. Recently, observers that do not depend on the
initial conditions or the estimated parameters from the standpoint
of asymptotic exponentially fast convergence to zero of the errors
have been built for many systems. They are called {\em globally
convergent adaptive observers} and have been obtained from a non
trivial combination of a nonlinear high gain observer and a linear
adaptive observer, see \cite{Zhang 2002} and \cite{Besancon 2004}.
In this work, we present an application of the high gain techniques
in the context of state estimation whatever the unknown parameter
is.

The assumption on the considered class of systems are basically that
if all of the parameters were known, some high-gain observer could
be designed in a classical way, and that the system are
``sufficiently excited" in a sense which is close to the usually
required assumption on adaptive systems, that is, signals should be
dynamically rich enough so that the unknown parameters can indeed be
identified. In this particular case, the lack of persistent
excitation of the system could impede the reconstruction of the
parameters. However, the recent scheme of Besan\c con and
collaborators \cite{Besancon 2004} guarantees the accurate
estimation of the states according to rigorous arguments in their
paper.

To make this mathematically precise we have to introduce first some
terminology. Let us construct the $j$th time derivative of the
output. This can be expressed using Lie differentiation of the
function $h$ by means of the vector field $f$ given by the right
hand sides of Norton's system. We will denote the $j$th Lie
derivative of $h$ with respect to $f$ by
${L_{f}}^{j}\left(h\right)\left(Z(t)\right)$. 
These Lie derivatives are defined inductively as functions of $Z$
\begin{eqnarray}
{L_{f}}^{0}\left(h\right)\left(Z\right) &=& h\left(Z\right)
\nonumber\\
{L_{f}}^{j}\left(h\right)\left(Z\right) &=& \frac{\partial}{\partial
Z}\left( {L_{f}}^{j-1}\left(h\right)\left(Z\right)\right)f(Z)~.
\nonumber
\end{eqnarray}
When the system is observable, i.e., from the knowledge of the
output one can build the states of the system, the Lie map $\Phi : Z
\rightarrow \Phi(Z)$ given by
\begin{eqnarray}
 \xi =\Phi(Z) = \left (
\begin{array}{c}
h(Z)\\
 {L_{f}}\left(h(Z)\right)
\end{array}
\right) = \left [
\begin{array}{c}
Z_{{1}}\\
K_{{1}}Z_{{1}}Z_{{2}}
\end{array}
\right ]
\end{eqnarray}
is a diffeomorphism. For $\Phi(Z)$ to be a diffeomorphism on a
region $\Omega$, it is necessary and sufficient that the Jacobian
$\mathrm{d}\Phi(Z)$ be nonsingular on $\Omega$ and that $\Phi(Z)$ be
one-to-one from $\Omega$ to $\Phi(\Omega)$, see \cite{Shim}.

Since $\Phi(Z)$ is a diffeomorphism, one can write the global
coordinate system defined by $X=\Phi(Z)$ in the following form

\begin{eqnarray}
\Upsilon: \ \ \ \left \{
\begin{array}{c}
\dot{X}_1 = X_2\\
\dot{X}_2 = \frac{{X_2}^2}{X_1}-K_2 X_2\\
 y = X_1
\end{array}
\right. ~. \nonumber
\end{eqnarray}

Following \cite{Besancon 2004}, we assume that the $\Upsilon$ system
can be written in the matrix form as follows



\newenvironment{mycase}{\left.\def\arraystretch{1.2} \array{@{}l@{\quad}l@{}}}{ \endarray \ \right\} }\makeatother

\begin{equation}
    \begin{mycase}
\dot{X} = AX+\varphi(X)+\Psi(X) \theta \nonumber\\
y = C X  \nonumber
 \end{mycase} \, \equiv 
 \digamma \,, \nonumber
\end{equation}

\noindent where $X \in \mathbb{R}^n$, $A=\left[
\begin{array}{cc}
0 & 1\\
0 & 0
\end{array}
\right]$, $\varphi(X)= (0, \frac{X_2^2}{X_1})^{T}$, $y$ is the
measured output, $\Psi(t) \in \mathbb{R}^{n \times p}$ is the matrix
of known functions and $\theta \in \mathbb{R}^p$ is the vector of
unknown parameters that should be estimated through the measurements
of the output $y$.
We are here in the particular case $n=2, \, p=1$, i.e.,
$\Psi(t)=[0,-X_2]^{T}$ and $\theta \in \mathbb{R}^1=K_2$. In
addition, the algorithm we develop is a particular case of that
presented in \cite{Besancon 2004}, since we can not meddle in the
system, in other words, there is no control input. As adaptive
observer we use the system \cite{Besancon 2004}
\begin{equation}
   \hat{\digamma}\equiv  \left \{ \begin{array}{l}
    \dot{\hat{X}} = A \hat{X}+ \varphi(\sigma(\hat{X})) + \Psi(\sigma({\hat{X}})) \sigma({\theta}) + \Lambda^{-1}
        \left [ \rho \mathcal{K} + \Gamma \Gamma^T C^T  \right]  \left(y  - C \hat{X}  \right) \\
    \dot{\hat{\theta}} = \left [\rho^n \Gamma^T C^T \right] \left(y  - C \hat{X}  \right) \\
    \dot{\Gamma}= \rho \left( A-\mathcal{K} C \right) \Gamma +\rho
    \Psi(\sigma(\hat{X}))
    \end{array}\right .
\end{equation}
\noindent where $\sigma(\cdot)$ is a saturation function, $\Gamma
\in \mathbb{R}^{n \times p}$ is the so-called gain vector,
$\mathcal{K}$ is a vector that makes $A-\mathcal{K}C$ a stable
matrix, $\Lambda = {\rm diag}[1, \rho^{-1}, \ldots, \rho^{-(n-1)}]$
where $\rho \in \mathbb{R}_{+}$ is a constant to be chosen. The
saturation function is a map whose image is bounded by chosen upper
and lower limits, $B$ and $b$, respectively. It is customary to
introduce such functions of simple forms, e.g., we used
\begin{equation}\label{sigma}
\sigma(s)=\\\ \left \{ \begin{array}{l} 
    B $\quad \quad$ s>B\\
    s $\quad \quad$ b\leq s \leq B\\
    b $\quad \quad$ s<b~,
    \end{array}
    \right . 
   \nonumber  
   \end{equation}
to avoid the over and/or underestimation and in this way to increase
the chance of the quick convergence to the true value \cite{Khalil}.

In \cite{Besancon 2004}, it is proven that the dynamical system
$\hat{\digamma}$ is a global exponential adaptive observer for the
system $\digamma$, i.e., for any initial conditions $X(t_0)$,
$\hat{X}(t_0)$, $\hat{\theta}(t_0)$ and $\forall \theta \in
\mathbb{R}^p$, the errors $\hat{X}(t)-X(t)$ and
$\hat{\theta}(t)-\theta(t)$ tend to zero exponentially fast when $t
\rightarrow \infty$. Taking
$\mathcal{K}=[\mathcal{K}_1,\mathcal{K}_2]$, the matrix $A-
\mathcal{K}C$ have the following eigenvalues
\begin{equation}
\lambda_{1,2} = -1/2\,\mathcal{K}_{{1}}\pm 1/2\,\sqrt
{{\mathcal{K}_{{1}}}^{2}-4\,\mathcal{K}_{{2}}}~.
\end{equation}
\noindent Selecting
$\mathcal{K}_{{2}}=(1/4)\,{\mathcal{K}_{{1}}}^{2}$, we get equal
eigenvalues $\lambda_1=\lambda_2= -(1/2)\,\mathcal{K}_{{1}}$, and
choosing $\mathcal{K}_1 >0$ we turn $A- \mathcal{K}C$ into a stable
matrix. Thus, the explicit form of the observer system
$\hat{\digamma}$ is given by

\begin{equation}\label{explicit}
\hat{\digamma}_{G}=\\\ \left \{ \begin{array}{l} \dot{\hat{X}}_1 =
\hat{X}_2 + \left( \rho \mathcal{K}_1 +
{\Gamma_1}^2\right) (X_1-\hat{X}_1) \nonumber\\
\dot{\hat{X}}_2 =
\frac{\left(\sigma(\hat{X}_2)\right)^2}{\sigma(\hat{X}_1)}-\sigma(\hat{X}_2)\sigma(\hat{\theta})+
\rho
\left( \frac{\rho {\mathcal{K}_1}^2}{4}+ \Gamma_1 \Gamma_2 \right)(X_1-\hat{X}_1) \nonumber\\
\dot{\hat {\theta} } = \rho^2 \Gamma_1 (X_1-\hat{X}_1) \nonumber\\
\dot{\Gamma}_1 = \rho (-\mathcal{K}_1 \Gamma_1 + \Gamma_2) \nonumber\\
\dot{\Gamma}_2 = - \frac{\rho {\mathcal{K}_1}^2 \Gamma_1} {4}+ \rho
\sigma(\hat{X}_2)~.\nonumber
\end{array} \right .
\end{equation}

Being global, this observer system does not depend on the initial
conditions. Therefore, any initial conditions chosen at random from
a set of physical values will not affect the correct estimation;
merely the convergence time could be longer or shorter. Thus, in
practice, it is useful to start with initial conditions that are
close to the real phenomenological initial conditions in a given
framework.

Finally, to recover the original states, we use the inverse
transformation $\Phi ^{-1}(\hat{X})$, which is given by:

\begin{eqnarray}
 \hat{Z}=\xi^{-1} =\Phi ^{-1}(\hat{X}) = \left (
\begin{array}{c}
\hat{Z}_1\\
\hat{ Z}_2
\end{array}
\right) = \left (
\begin{array}{c}
\hat{X}_{{1}}\\
\frac{\hat{X}_{{2}}^2}{K_{{1}}\hat{X}_{{1}}}
\end{array}
\right )~.
\end{eqnarray}

\begin{figure}[x]
\centerline{
\includegraphics[scale=.7]{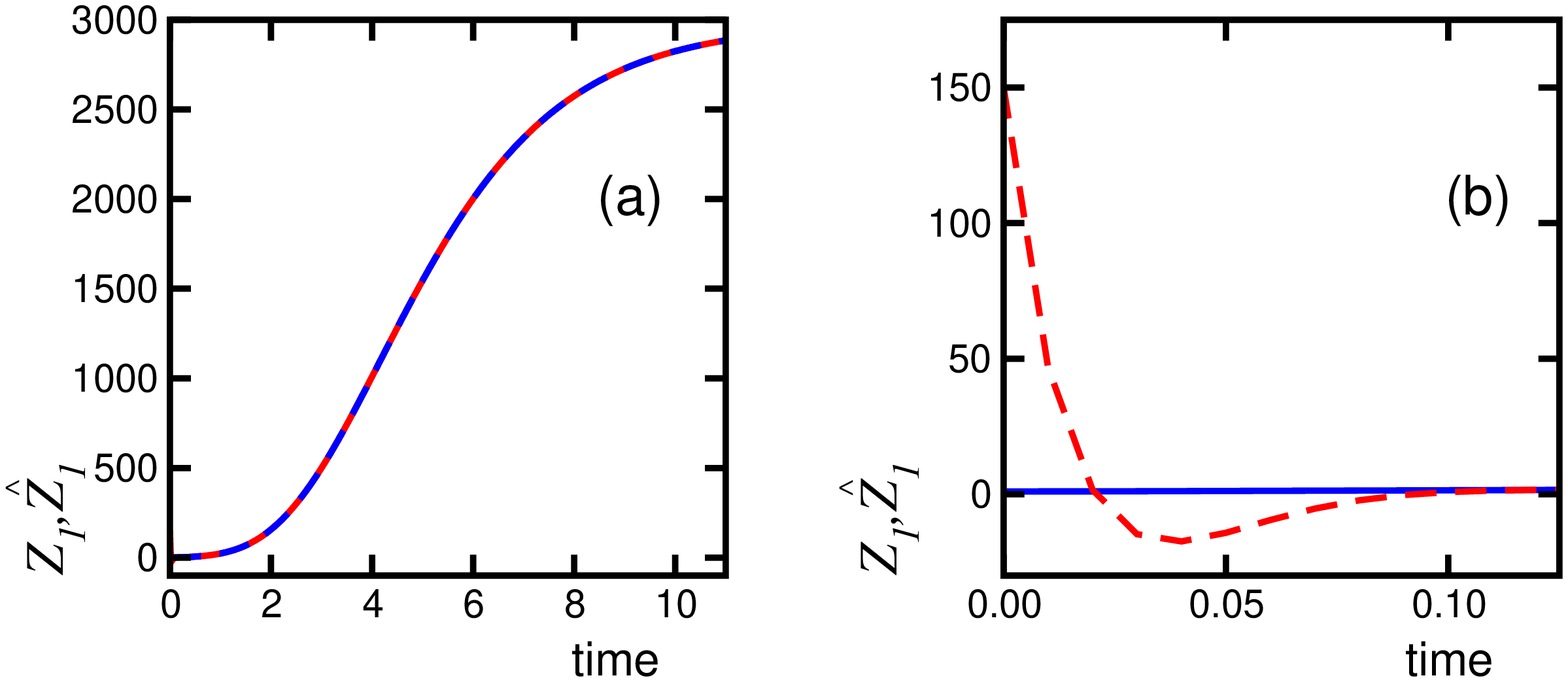}}
  \caption{ Numerical simulation for the first state: (a) the solid line represents the time evolution of the true states $Z_1$ and the dotted
 line represents the estimate $\hat{Z}_1$. Plot (b) is a detail of (a) to appreciate the variation of
   $\hat{Z}_1$ in the beginning.} \label{simulaciones X1}
\end{figure}
\begin{figure}[x]
\centerline{
\includegraphics[scale=.7]{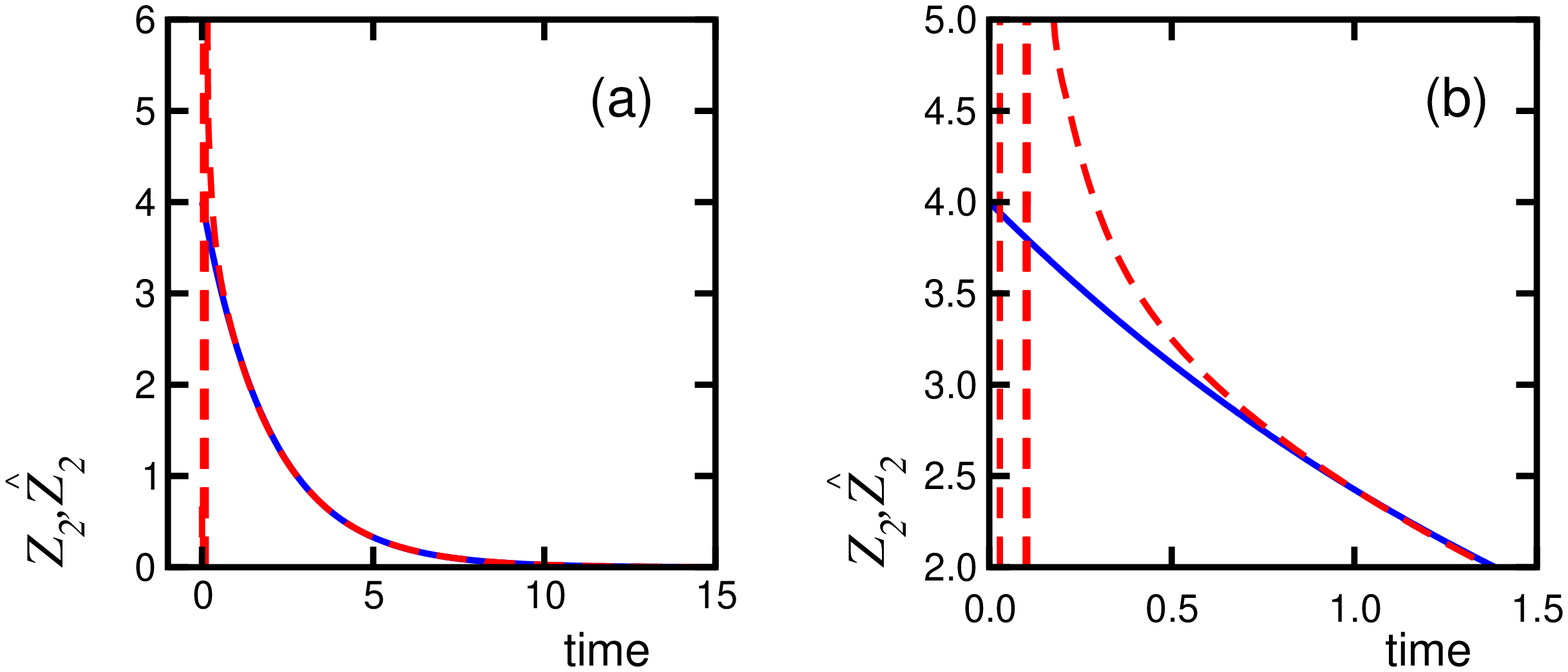}}

  \caption{Numerical simulation for the second state: (a) the solid line represents the true state $Z_2$ and the dotted
  line represents the estimate $\hat{Z}_2$. Plot (b) is a detail of (a) in order to appreciate better the variation of
   $\hat{Z}_2$ in the beginning.}\label{simulaciones X2}
\end{figure}

With the aim of better illustrating the adaptive scheme proposed
here, we present numerical simulations. We use the following values
of the parameters: $K_1=1$, $K_2=0.5$, $\rho=100$ and
$\mathcal{K}_1=1$. In Figs. (\ref{simulaciones X1}) and
(\ref{simulaciones X2}), the solid lines represent the evolution of
the true states and the dotted lines stand for the evolution of the
estimates, respectively. We mention that short convergence time is
what really matters in order to have efficient numerical
simulations. This can be accomplished by starting with arbitrary
initial conditions that are guessed to be close to the real initial
ones as we already commented. If one is interested in the evolution
of the iterative scheme, this can be readily glimpsed from the
difference $\hat{Z}_i-Z_i$ between the curves in the figures.

To this end, we would like to illustrate the robustness of the
present adaptive scheme. Figs.~(3) and (4) show what happens when an
impulsive type perturbation (i.e., of high value acting in a very
short span of time) is added to the output signal $h(Z)=Z_1$ which
is fed to the observer at $t=4$ (arbitrary units). This perturbation
is a strong error in the output $y$ that can be attributed to an
unexpected fault in a measurement device or even human errors and
does not belong to the natural evolution of the tumour. As can be
seen from the graphics, the adaptive scheme has the ability to
recover the ``true" signal immediately after the perturbation
disappears. In general, this robustness is due to the fact that the
scheme is designed in the closed-loop way and additionally not the
full range of the parameters need to be known.

\begin{figure}[x]
 \includegraphics[scale=.7]{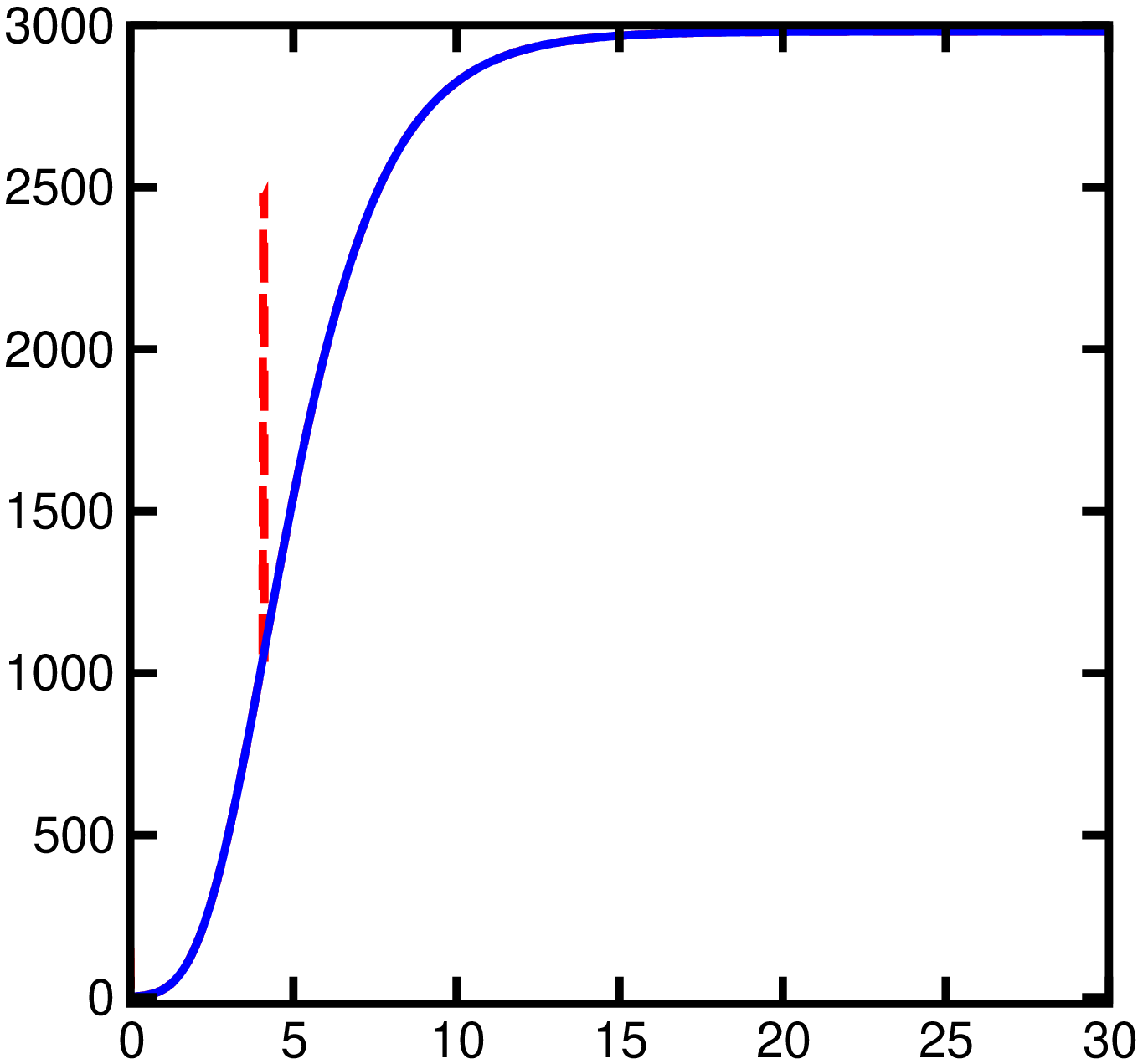}  
\caption{Behaviour of $Z_1$, $\hat{Z}_1$ under an impulse-type
perturbation at $t=4$ (arbitrary units).}
\end{figure}


\begin{figure}[h]
 \includegraphics[scale=.7]{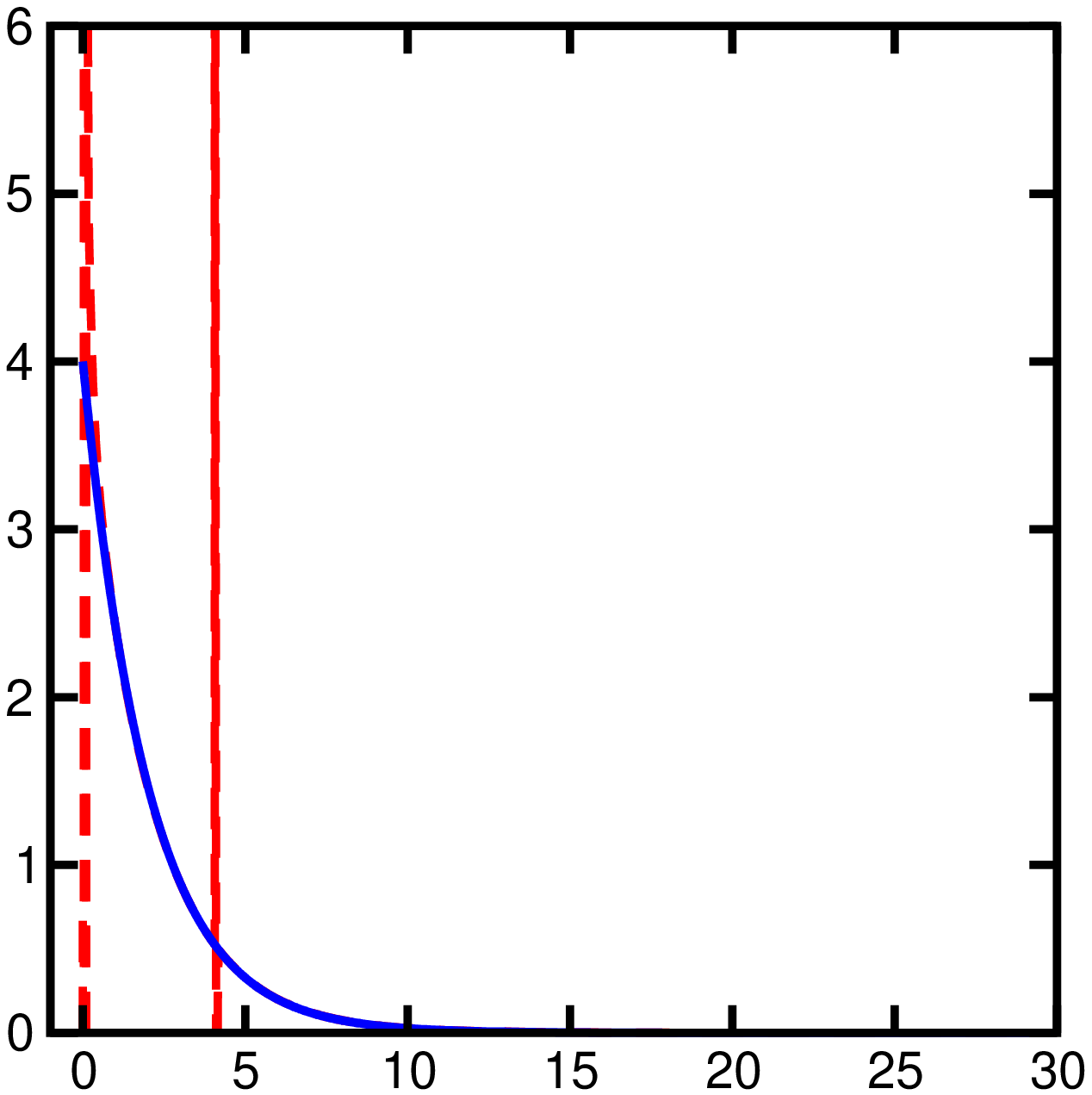} 
\caption{Behaviour of $Z_2$, $\hat{Z}_2$ under the same perturbation
at $t=4$ (arbitrary units).}
\end{figure}

\section{Conclusion}

The robust adaptive scheme we used here for the interesting case of
Gompertz growth functions is a version of that due to Besan\c con
{\em et al}. The results of this work indicate that this scheme is
very efficient in obtaining the Gompertz functions without knowing
both initial conditions and parameter $K_2$. The method may be
useful in more general frameworks for models of self-limited growth
such as in the construction of a specific growth curve in biology,
or as a managerial tool in livestock enterprizes, as well as in the
detailed understanding of the growth of tumors. We also notice that
the reconstruction of the unknown states by this method allows the
possibility to obtain important missing parameters by standard
fitting procedures.


\end{document}